\documentclass[12pt, letterpaper]{article}

\usepackage{amsmath,amssymb}
\usepackage{cite}
\usepackage{fancyhdr}
\usepackage[top=1in, bottom=1in, left=1in, right=1in]{geometry}
\usepackage[dvipdfm, dvips]{graphicx}
\usepackage{hyperref}

\numberwithin{equation}{section}

\begin{document}


\setcounter{page}{0}
\date{}

\lhead{}\chead{}\rhead{\footnotesize{RUNHETC-2010-17\\SCIPP-10-12}}\lfoot{}\cfoot{}\rfoot{}

\title{\textbf{Pedagogical notes on Black Holes, de Sitter space and bifurcated horizons\vspace{0.4cm}}}

\author{Tom Banks$^{1,2}$\vspace{0.7cm}\\
{\normalsize{$^1$NHETC and Department of Physics and Astronomy, Rutgers University,}}\\
{\normalsize{Piscataway, NJ 08854-8019, USA}}\vspace{0.2cm}\\
{\normalsize{$^2$SCIPP and Department of Physics, University of California,}}\\
{\normalsize{Santa Cruz, CA 95064-1077, USA}}}

\maketitle
\thispagestyle{fancy}

\begin{abstract}
\normalsize \noindent I discuss black hole evaporation in two
different coordinate systems and argue that the results of the two
are compatible once one takes the holographic principle into
account. de Sitter space is then discussed along similar lines.
Finally I make some remarks about smooth initial conditions in GR,
which evolve to space-times with bifurcate horizons, and emphasize
the care one must take in identifying spaces of solutions of General
Relativity which belong to the same quantum theory of gravity. No
really new material is presented, but the point of view I take on
all 3 subjects is not widely appreciated.

\end{abstract}


\newpage
\tableofcontents
\vspace{1cm}


\section{Black hole evaporation}

The string theoretic arguments for unitarity in black hole
evaporation have been accepted by many researchers and dismissed by
others. Even those who accept them tend to feel unease about the
absence of a more local picture of how Hawking's paradox is
resolved.  Some time ago, W. Fischler and I \cite{tbwf} proposed
such a resolution, but it has not garnered much attention.  It seems
to me that by simply adding a little bit of information, namely the
holographic principle, about the non-perturbative formulation of
quantum gravity, one comes to a satisfactory resolution of the
paradox without all of the heavy machinery of string theory.  The
demonstration proceeds by tackling the singular region of the
space-time geometry head on, and showing that the holographic
principle\footnote{Throughout this note I will be referring to the
Strong form of the Holographic Principle advocated by Fischler and
myself. This is the assumption that the density matrix referred to
in the covariant entropy bound is the maximally uncertain density
matrix on the Hilbert space associated with the region bounded by
the holographic screen.} provides a novel interpretation of it,
which ends up being completely equivalent to the Schwarzschild
observer's point of view.

This equivalence will also motivate some remarks about the nature of
the states in the two pictures of the geometry, which are even more
important for appreciating the nature of the quantum theory of de
Sitter space.

In Schwarzschild coordinates, the eternal black hole metric is

$$ds^2 = - dt^2 f(r) + \frac{dr^2}{f(r)} + r^2 d\Omega^2, $$ where
$f(r) = (1 - \frac{R_S}{r})$, and $d\Omega^2$ is the metric on the
unit two-sphere. Near the horizon, where $r = R_S + x$, this becomes
$$ ds^2 = - dt^2 \frac{x}{R_S} + x R_S (\frac{dx}{x})^2 + R_S^2
d \Omega^2 .$$ This is the same as
$$ ds^2 = R_S^2[ e^y (- dt^2 + dy^2) + d\Omega^2 ],$$ where we have
made all coordinates dimensionless and committed the unforgivable
sin of using the same letter for dimensionless and dimension-full
time.  The near horizon limit is $y \rightarrow - \infty$.

As a consequence, states with arbitrary energy according to a local
geodesic observer, have arbitrarily small energy as measured by the
supported observer at infinity.  This is the famous redshift of
signals coming from an observer falling into a black hole.  The
supported observer never sees anything fall through the horizon, but
signals from things near the horizon become harder and harder to
detect. Equally important is that the range of the $y$ coordinate
near the horizon is infinite, and that the metric there is conformal
to flat space. When we use the boundary conditions at the horizon
which correspond to the Hartle-Hawking\cite{hh} vacuum state on the
extended Kruskal manifold, quantum fields in this geometry have an
infinite number of arbitrarily low energy states, all concentrated
near the horizon.

This contradicts the Bekenstein-Hawking formula, which assigns a
finite entropy to the black hole. The holographic principle states,
in this context, simply that the BH formula is correct, and that
quantum field theory over-counts the near horizon states by an
infinite amount.

Now let us introduce another coordinate system.  Consider the $t=0$
slice of the Schwarzschild coordinates and time-like geodesics that
penetrate this slice orthogonal to it. Follow along each of these
geodesics, measuring time according to the local proper time. Label
each point on the new time slices by the same coordinates that label
the intersection of the same geodesic with the $t = \tau = 0$ slice.
It is important, and true for the Schwarzschild metric, that the
time-like geodesics never cross.  This gives us a metric of the form
$$ds^2 = - d\tau^2 + R_S^2 \frac{(R^2 + 1)}{R^2}  (\frac{d r}{d R})^2 dR^2 +
 r^2 (R,\tau) d\Omega^2 . $$
The Schwarzschild radial coordinate $r$ is defined as a function of
$\tau$ and $R$ by
$$ \frac{\tau}{R_S}= \pm (R^2 + 1) \bigl[ \frac{r}{R_S} - \frac{r^2}{(R_S^2 (R^2 + 1))}\bigr]^{1/2}$$
$$ +  (R^2 + 1)^{3/2} \cos^{-1}\bigl( [\frac{r}{R_S (R^2 + 1)}]^{1/2} \bigr) .$$
These are called Novikov coordinates\cite{mtw}, and they cover the
entire Kruskal manifold just as well as Kruskal coordinates.  They
also match to the Schwarzschild coordinates at infinity.  The
metric, as well as the Hamiltonian of quantum fields in this
coordinate system, is time dependent.  Inside the horizon the time
dependence becomes singular at a finite proper time of order $R_S$.
This is the surface of Schwarzschild coordinate $r=0$. We expect
more and more field theoretic degrees of freedom to be excited as we
approach the singularity.

It is important to note that any finite spatial coordinate $R$ maps
to a point inside the horizon at large enough $\tau$.  The Novikov
coordinates map out a congruence of time-like geodesics, each of
which starts with zero velocity at $t = \tau = 0$, and some value of
$R$ and the angles on the two sphere.  Any such geodesic is
attracted by the black hole and falls through the horizon and
encounters the singularity. For a given $R$, the time it takes to
hit the singularity is $\sim R$.  However, it is only for the last
$\sim R_S$ of that time, that this coordinate point is inside the
horizon.   Thus, at a given time $\tau$ there is a value $R(\tau )$,
such that all $R > R(\tau )$ are outside the horizon.  Below we will
discuss time scales of order $\tau \sim R_S$ and use the phrase {\it
inside the horizon} to mean ``those values of $R$ which are inside
the horizon at this time".   Quantum field theory can be split into
degrees of freedom localized in the interior and exterior regions,
and in this coordinate system there is no funny buildup of states on
the horizon.  We will now argue that the QFT description is not
valid, in these coordinates, in the interior.

It is a consequence of the singularity that no causal diamond inside
the horizon has a holographic screen area larger than the area of
the black hole horizon.  The strong holographic principle then
implies that the system inside the horizon has a finite number of
states, contrary to the prediction of quantum field theory.

On the time scale $R_S$ the geometrical picture of the spatial
geometry of the interior of the black hole, in Novikov coordinates,
breaks down.  The black hole singularity $r=0$ is a space-like
surface that occurs at fixed Novikov time of order $R_S$. To
understand the spatial geometry, suppress one coordinate on the two
sphere, and think of the standard picture of the geometry of a
static star as a bowl shaped depression, with the matter in the star
at the bottom of the bowl. In spherical gravitational collapse, the
bowl gets deeper and its cross section gets narrower, reaching
infinite depth and zero width in a time of order $R_S$.

Consider a collection of ``sputniks", thrown into the black hole to
map out its interior geometry. We can view these as a physical
construction of Novikov coordinates in the interior. If two of these
satellites are thrown into the black hole with a small time
difference, they are out of causal contact. The space between them
stretches faster than the speed of light. If, as in the synchronous
coordinates, they are thrown in at precisely the same time, they
will instead crash into each other in a time no bigger than $R_S$,
if they have any width at all. More mathematically, it is well known
that small classical anisotropic fluctuations grow rapidly as the
singularity is approached, and that there is rapid quantum particle
creation for quantum fields placed in this background.  The obvious
conclusion is that, in Novikov coordinates, the whole quantum field
theory in a fixed space-time background approximation is breaking
down rapidly as the singularity is approached.   For $R_S \sim
10^{-9} {\rm sec} \sim 10 {\rm cm},$ the black hole evaporation time
is of order the age of the observed universe.  Thus, in Novikov
coordinates, the singularity is encountered long before the external
observer can see enough Hawking radiation to make {\it any}
conclusion about the internal state of the black hole.

The time dependent Hamiltonian in Novikov coordinates inside the
horizon, is not integrable.   It commutes with the space-like
Schwarzschild Killing vector $\partial_t$, and with the generators
of $SO(3)$, but has no other conservation laws.   This fact, and the
holographic principle, are all we need to understand the behavior of
the system inside the horizon on times scales of order $R_S$. There
is no meaning to the interior time evolution for larger times. In
this regime, it is a good approximation to treat the interior of the
black hole as an isolated system. Hawking radiation is transferring
degrees of freedom to the external world, but for large $R_S$, this
process is slow and its effect on the interior dynamics can be
neglected.

Consider the subspace of the interior Hilbert space with fixed
values of the conserved quantities.  The fact that the system is not
integrable is equivalent to the statement that the Lie algebra
generated by the time dependent Hamiltonians $H(\tau )$ is the
algebra of all Hermitian matrices on this subspace.  We will also
assume that this remains true near the singularity at $\tau = \tau_s
= k R_S$.  An example of a way in which this could fail is a
behavior like  $$H(\tau ) \sim H_s (\tau - \tau_s)^{-p} .$$  In this
case the behavior near the singularity would become integrable and
the eigenstates of $H_s$ would be preserved by the evolution, up to
a rapidly oscillating phase\footnote{If the number of states is
large, and the Hamiltonian $H_s$ sufficiently generic, then this
integrability is illusory, and the conclusion we will draw is still
valid after time averaging.}. There is no indication of this kind of
behavior in the field theory Hamiltonians in the black hole
background. Indeed, in field theory the Hilbert space can be viewed
in the the Fock space basis of in-falling particles. As the
singularity is approached we can excite more and more particles of
higher and higher energy, and the state vector wanders off into the
highest entropy part of the Fock space. Of course, in field theory,
the Hilbert space is infinite dimensional, so the evolution runs off
to infinity and the system is truly singular.

By contrast, if we accept the bound on the number of states implied
by the Strong Holographic Principle, the space of normalizable
states is a compact manifold, $CP^N$. The one parameter
non-integrable groupoid of unitary transformations $T e^{ -i \int^t\
ds\ H(s)} $ describes a chaotic trajectory on this manifold. The
singularity in the flow implies that each point in $CP^N$ is visited
an infinite number of times as the singularity is approached. The
time averaged state of the system rapidly approaches the maximally
uncertain density matrix on the finite dimensional space of black
hole interior states.  Although the QFT in curved space-time
description of the dynamics has completely broken down, very simple
general principles allow us to conclude what the time averaged
behavior of the system is, with no need to find a precise
description of the interior Hamiltonian.

Indeed, from a more ambitious point of view, we may say that there
is no meaning to the effective Hamiltonian for this localized
region, besides the general behavior we have described. The
Principle of General Covariance tells us that we should be able to
describe the interior of the black hole with many different
coordinate systems. A small sample of those would be systems of
synchronous coordinates which differ from Novikov coordinates only
by the choice of the portion of the initial space-like slice inside
the horizon.  The Hamiltonians $H(\tau )$ in this new set of
coordinates will not commute with those in Novikov coordinates, but
will have the same general properties.  The detailed time dependence
of the state will differ, but the time averaged behavior will be
identical. Thus, one could view the conclusion that the time
averaged density matrix of the black hole interior rapidly
approaches the maximally uncertain density matrix on a Hilbert space
of dimension determined by the strong holographic principle, as the
only generally covariant fact about the black hole interior.

In any of these {\it Novikovoid} coordinate systems one can
describe, in the QFT approximation, the Hilbert space of the region
outside the black hole horizon. For large $R_S$ the Fock space basis
of multi-particle states is the best basis in which to think about
the time evolution.  The Arnowitt-Deser-Misner\cite{adm}
Hamiltonian, which is defined as a surface term at infinity, acts on
this exterior Hilbert space. The holographic principle tells us that
we should think of this particle system as interacting with another
finite dimensional quantum system. On time scales of order $R_S$,
the discussion of the previous paragraphs tell us that this system
approaches a time averaged state which is the maximally uncertain
finite dimensional density matrix.  All of these states have the
same ADM energy, so it makes sense to say that the black hole
interior is in its micro-canonical ensemble at energy equal to the
black hole mass.  Since its entropy, $\pi (R_S M_P)^2$, is large,
the interaction of field theory with this system should be well
described by thermodynamics.

The rest is history.  We have discovered that, given very generic
properties of the holographic description of the interior, the
physics predicted for the exterior on a time scale greater than
$R_S$ and less than $R_S^3 M_P^2 $ is identical to that in
Schwarzschild coordinates. The fearsome transition between pure and
mixed states has been accomplished by the traditional pragmatic
method of time averaging the state of an ergodic system.  On time
scales of order the black hole evaporation time, one would need a
more sophisticated model which allows the exchange of degrees of
freedom between the interior and exterior systems.  The only models
I know which come anywhere near achieving this goal are those of
\cite{matbh}

There are a number of important lessons we have learned along the
way.  Perhaps the most important is that we should not expect to
find a detailed microscopic quantum description of the interior of a
black hole.  In general relativity, local physics is gauge
dependent. When a local region is well described in terms of quantum
particles traveling through a classical space-time, the notion of a
gauge independent local description is approximately meaningful.
Within the black hole interior, there are no semi-classical
observables to catch hold of.  No basis of the Hilbert space is
preferred and the only invariant conclusion we can come to is that
the density matrix of the system is maximally uncertain.

Next we emphasize the fact that the black hole singularity has been
``resolved" by quantum mechanics, without removing it. Indeed the
singular behavior of the time dependent finite dimensional
Hamiltonian is crucial to the conclusion that the density matrix
rapidly approaches the micro-canonical ensemble as the singularity
is approached.  This is quite different from the resolution of
time-like singularities by the addition of IR quantum degrees of
freedom, which has been the hallmark of the string theory treatment
of singularities.

Another important feature of our discussion has been the coincidence
between the description of the gauge invariant physics in the
Schwarzschild and Novikov coordinate systems. The misleading
infinity in the state count is encountered in both systems, and is
cut off by fiat, using the holographic principle. The description of
the physics of black hole interior states (which the Schwarzschild
observer calls near horizon states) is different because the set of
observables used to probe them in these different coordinate systems
are different, and do not commute.  However, in both coordinate
systems we conclude that the state of these degrees of freedom is
well approximated by a high entropy density matrix. More detailed
quantum information about these states is really available only to
the external observer.  Ultimately what this means is that there is
an S-matrix in asymptotically flat space, which completely describes
the formation and evaporation of black holes.  It is unlikely that
anyone will ever find even an approximate description of that
S-matrix, which goes beyond the inclusive information first
calculated by Hawking.  For the truly interior description, we have
argued that the detailed microscopic description of the dynamics
appears in the semi-classical approximation to depend strongly on
the choice of interior time slicing.  Since there is no operational
way to actually construct these coordinate systems, those
descriptions seem somewhat meaningless, except that one finds the
same time averaged density matrix near the singularity, for all of
them.  The Schwarzschild description of interior states as near
horizon states, leads immediately to the conclusion that black hole
evaporation is a unitary process. The holographic principle allows
us to come to an identical conclusion in coordinate systems that are
smooth at the horizon.

In the early 1990s there was a renewed burst of activity related to
the paradoxes of Hawking radiation.  I believe that it was about
this time that the idea of a ``nice slice" coordinate system was
first realized to be crucial to Hawking's argument that black hole
evaporation violates unitarity. The nice slice coordinates are
similar to Novikov coordinates, in that they are smooth at the
horizon, but the interior time coordinate is related to the exterior
one on the same time slice by a large Lorentz boost. When time of
order the evaporation time $R_S^3 M_P^2$ has passed in the exterior,
the interior time is still significantly less than $R_S$.  This
enables one to claim that it is reasonable to use QFT to describe
the whole slice.  One then claims that information that an advocate
of unitarity would insist is encoded in the outgoing radiation, is
in fact encoded in field degrees of freedom in the interior portion
of the slice, which commute with all the exterior radiation fields.

There have been significant disagreements in the literature, even
among authors of the same paper\cite{joeetal}, about whether the
sub-Planckian time intervals between slices in the interior,
invalidates the use of quantum field theory in these coordinates. I
will not add more verbiage to this controversy, but simply claim
that our discussion shows that one can use less contrived coordinate
systems, all smooth at the horizon, to come to conclusions opposite
to those indicated by doing field theory on nice slices.  Like many
other overly contrived confections, nice slices are profoundly
misleading. Given our discussion, one can no longer state that the
Schwarzschild coordinate view of black hole evaporation is
misleading because it uses coordinates that are singular at the
horizon.

I want to end this section by emphasizing again that our discussion
is only approximate and does not extend to a description of the full
process of Hawking evaporation of a black hole.  In order to do that
one would have to find a method in which one could allow a finite
quantum system interacting with a QFT to ``sublime" and transmit
almost all of its degrees of freedom into outgoing radiation. It's
clear that if this can be done at all, the proper description of the
finite system would be a cutoff version of the near horizon states
in Schwarzschild coordinates.  The problem that bedevils any such
attempt is that in order to describe the system properly, one has to
let the background geometry change in order to conserve ADM energy.
I suspect that the only complete description of the evaporation
process is through the S-matrix of a model of quantum gravity in
asymptotically flat space.  From the practical point of view we
should admit that computing that S-matrix should be at least as
difficult as a complete S-matrix description of the byproducts of a
thermonuclear explosion starting from a description of the initial
state of some colliding nuclei.

A more local description of black hole physics might be achievable
with the formalism of holographic space-time\cite{tbwf2}.  There the
fundamental object is the quantum version of a causal diamond with
finite area holographic screen.  This is just a finite dimensional
Hilbert space, which for large dimension encodes the area of the
screen through the Bekenstein-Hawking relation. The intersection
between two diamonds is described by a common tensor factor in their
Hilbert spaces. A holographic space-time is a collection of Hilbert
spaces, with overlap prescriptions, and a consistent unitary
dynamics. One can encode local concepts approximately in this
formalism. For example, a time-like trajectory is encoded as a
nested sequence of Hilbert spaces, each a tensor factor of its
successor.

For a black hole, consider a causal diamond $I$,which has its past
tips prior to the black hole formation. $I$ lies mostly inside the
horizon, but its outer boundary sticks out of the horizon over a
range of solid angles, centered at some solid angle
$\overrightarrow{n}$. Now consider a causal diamond $E$, which lies
mostly outside the horizon, but has part of its boundary dipping
down into the horizon and incorporating the portion of the boundary
of $I$ outside the horizon.  Now consider a collection of such pairs
of diamonds tiling the sphere.  That is, the intersection of any
pair of diamonds $E_i$ and $E_j$ is zero, and any point on the
horizon is contained in some $E_i$.  We think of the collection of
$E_i$ as the quantum system corresponding to the ``stretched
horizon".

I believe that it is possible to prove that the areas of the
holographic screens of the maximal diamonds contained in the
overlaps between $I_i$ and $E_i$ is $\geq$ the area of the horizon,
because all 2-surfaces inside the horizon are trapped.  If one
relies on the geometrical/QFT picture of the black hole interior,
one would argue that there is important information in $I_i$ that is
causally disconnected from the overlap.  Indeed, one can think of
$E_i$ as the causal diamond of an observer that is supported near
the horizon for a long time, and eventually falls into the black
hole. The future tip of $E_i$ must be on the singularity, but there
are timelike trajectories in $E_i$ that can stay outside the horizon
for a very long time\footnote{It is always worth keeping in mind
that the black hole doesn't really last forever.  One can easily
choose $E$ so that it contains timelike trajectories which stay
outside the horizon for times of order the evaporation time.}. Such
an observer is causally disconnected from observers that fell into
the hole much earlier.  However, our discussion has shown that for
times of order $R_S$ in Novikov coordinates, one should not apply
the QFT picture to the interior of the black hole. Instead it is a
finite dimensional quantum system, which revisits every one of its
states an infinite number of times as $\tau$ approaches the
singularity.  The degrees of freedom in $I_i$ may be only a tensor
factor in that interior Hilbert space, but the collection of all
$I_i$ encompasses all of it.

In the holographic space-time approach, there is room in the
collection of $E_i$ to encode all of the information in the
collection of $I_i$.  We have argued that the dimension of the
tensor product of these Hilbert spaces is large enough to teleport
the state of the interior into a state in $\otimes_i E_i$.
Furthermore, the rules of the holographic formalism say that $E_i$
contains a tensor factor identical to $I_i$, and whose state is a
unitary transform of the state in $I_i$ (part of the definition of
the quantum space-time is the specification of this unitary
transformation).  We have had to give up a quasi-local description
of $I_i$ in order to achieve this teleportation, but that is all.
There are larger causal diamonds, describing the trajectories of
observers that remain outside the black hole forever, which
intersect with the $E_i$ and can transmit all of the information
about the black hole interior, out to infinity. This process is
perfectly compatible with an approximate local field theory
description.

Thus, following the prescriptions of holographic space-time, and
prescribing that the Hilbert spaces describing the black hole
interior have a singular, ergodic, time dependent Hamiltonian, of
the type described above we get a picture of black hole evaporation
with only two lacunae.  The first is that there is not yet a
derivation of the dynamics of QFT in curved space-time as an
approximation to a consistent holographic space-time.  We have
identified the correct kinematic variables to describe particle
physics\cite{bfm} but do not yet have a consistent holographic
dynamics, which is close to that prescribed by QFT.

The holographic formalism may provide a way to get around the second
lacuna, the peculiar phenomenon of sublimation of degrees of
freedom, which we referred to above. In our heuristic discussion of
QFT in Rindler coordinates, we encountered different variables of
the system, whose time dependence became singular at different
Novikov times.  In the holographic description, the singular
Hamiltonian acts only on the Hilbert space of internal states.  The
external Hilbert spaces $E_i$ use a different time variable to
describe the dynamics of the black hole state, but agree with the
conclusion that its time averaged density matrix is micro-canonical.

The inner boundary of the causal diamond described by $E_i$, but its
outer boundary can reach null infinity.  Thus, there is no barrier,
in principle, to construct a dynamics in which a subset of the
degrees of freedom describing $E_i$ are in equilibrium for a long
time, but then turn into radiation.  In the holographic description,
different causal diamonds are independent quantum systems, with
their dynamics constrained to agree, up to conjugation by a unitary,
on overlaps. Causal diamonds of some in-falling observers have
overlaps with those of some supported observers, which consist of a
region close to the horizon.  The consistency conditions are
satisfied if the two descriptions both conclude that the finite
dimensional system describing the overlap, has a time averaged
density matrix that is maximally uncertain.

\section{de Sitter space}

I will keep this section brief, since I have discussed this material
in many recent articles.  We will examine two interesting coordinate
systems for dS space.  The first, called static coordinates, is the
analog of the Schwarzschild coordinate system for a black hole.  It
has the form
$$ds^2 = - dt^2 f(r) + \frac{dr^2}{f(r)} + r^2 d\Omega^2 ,$$ with
$$f(r) = (1 - \frac{r^2}{R^2}) .$$

Near the horizon $f(r)$ has a linear zero, and we can transform to
coordinates like $(t,y)$ for the black hole.  The Killing vector
$\partial_t $, which has norm $1$ at $r =0$ becomes null at the
horizon, so we again see an infinite pileup near the horizon of very
low energy states of QFT on this background.  As before, the
holographic principle suggests that this infinity is cut off, and
that the total entropy of the static patch is $\pi (RM_P)^2$ .

In global coordinates, the metric has the form

$$ds^2 = - d\tau^2 + R^2 \cosh^2 (\tau /R) d\Omega_3^2 ,$$ where
$d\Omega_3^2$ is the metric on the unit 3-sphere. Here the infinity
shows up as the infinite spatial volume of the 3-spheres as $|t|
\rightarrow\infty $.  At late times, it looks like we get an
infinite number of copies of the spatial region covered by the
static coordinates.  Note however that at $\tau = 0$ there are only
two copies, and as we will see in the next section, the second copy
is just a trick for computing thermal Green's functions in the
static region.  Since quantum field theory is unitary, there are no
more quantum states in the large $\tau$ region than there are at
$\tau = 0$.  In field theory, this is because there are an infinite
number of states in any volume, no matter how small. The expansion
of the universe can convert field theory states of any large
momentum, into low momentum states.

As in asymptotically flat or AdS spaces, we can obtain useful
information about the quantum theory by investigating perturbations
that do not disturb the asymptotic behavior.  Since most ways of
foliating this geometry give compact spatial sections, the
asymptotic regions to be considered are past and future infinity.

To get an idea of the constraints on such perturbations, consider
the exercise of setting small masses $m$ on each point of the
sphere, {\it i.e.} making the ``co-moving observers" of global
coordinates into physical particles. If we do this at global time
$T$, and space the masses by the particle's Compton wavelength
(since in a quantum theory, no particle can be localized more
precisely than that), then at $t= 0$ the particle number density is
$$m^3 \cosh^{3} (T/R), $$ and the $00$ component of the
stress tensor is exponentially large if $T \gg R$.  In other words,
long before $t = 0$, the back reaction on the geometry of the test
masses becomes important. In order to avoid this, we must make $m
\sim \cosh^{-1} (T/R)$ at time $T$.  This strongly suggests that,
{\it if we want to preserve dS asympotics in the future, we must not
try to fill the apparently huge volumes of space available in the
past with matter}.  Rigorous results along these lines have been
obtained in \cite{gary}\cite{boussofreivogel}.  The conclusion of
those studies is that if one inserts too much matter in the infinite
past, then a singularity forms before $t = 0$. If the singularity
can be confined within a marginally trapped surface of radius $<
3^{- \frac{1}{2}} R$, this can be viewed as a black hole excitation
of dS space, but if not, the whole space-time experiences a Big
Crunch and we are no longer within the class of asymptotically dS
space-times.

Thus considerations strictly confined to global coordinates give us
similar constraints on the total entropy of asymptotically dS
space-time, as those we find in the static patch. As in the black
hole example, the infinite number of horizon volumes in global
coordinates is mirrored by the infinite number of near horizon
states in static coordinates.  Both infinities are cut off by the
holographic principle.

We can get a little more insight into this by thinking about how
many of the states in the static patch can be thought of as
localized particles in the bulk.  The field theory entropy is of
order

$$S \sim M_c^3 R^3 ,$$ where $M_c$ is an ultraviolet cutoff. The
energy of a typical state in this ensemble is $$E \sim M_c^4 R^3,$$
and it has a Schwarzschild radius
$$R_S \sim \frac{M_c^4 R^3}{M_P^2} ,$$ so if we insist that the
system does not contain black holes whose size scales like $R$
(since most black hole states are not well modeled by field theory)
then we find the field theoretic entropy is bounded by
$$S < (RM_P )^{3/2} .$$  This indicates that the total dS entropy is
large enough to accommodate $(RM_P )^{1/2}$ copies of the maximal
set of field theory degrees of freedom in a single horizon volume.
This is, plausibly, the correct cutoff on the global coordinate
picture of an infinite number of horizon volumes in unconstrained
QFT.

To summarize, simple arguments, in either global or static
coordinates, suggest that dS space is a quantum system with a finite
dimensional Hilbert space, in accordance with the Strong Holographic
Principle.

\section{Bifurcate horizons, entanglement and density matrices}

Both the eternal black hole and de Sitter space have bifurcate
horizons.  The space-time geometry consists of two causally
disconnected regions separated by a pair of horizons. There is a
space-like surface on which the two regions touch at a single point,
so that, in QFT the Hilbert space can be thought of as a tensor
product\footnote{Actually, there are technical subtleties in this
statement having to do with UV divergences.  We will ignore those in
the sequel, primarily because a truly quantized theory of
gravitation cuts them off.} of Hilbert spaces of states localized in
the two separate regions.

W. Israel\cite{israel}, following the seminal work of Hartle and
Hawking, first pointed out the proper interpretation of this
peculiar feature of these geometries in the quantum theory. Both the
eternal black hole and dS space are thermal equilibrium states. The
QFT physics in such a state is contained in the thermal correlators

$${\rm Tr\ } e^{-\beta H} T \phi (x_1, t_1) \ldots \phi (x_n, t_n )
. $$

For equal times, the static thermal correlators are easily computed
in terms of a Euclidean path integral. In principle, more general
correlators can be obtained by analytic continuation, but analytic
continuation of an approximate formula often misses crucial aspects
of the real time physics, and in particular hydrodynamic behavior,
in which the results of small perturbations acting over long time
periods can be crucial. Instead, the real time physics is usually
represented by the Keldysh-Mahanttapa-Schwinger closed time path
formalism. This is a Lorentzian signature path integral over a
complex contour.  It computes in-in expectation values, unlike the
standard Lorentzian path integral, which produces in-out transition
amplitudes. By making insertions on different portions of the
complex contour, one can compute time ordered expectation values, or
anti-time ordered ones, or retarded commutators, or Wightman
functions {\it etc.}.

A particular form of the closed time path integral, called
Thermo-field theory, computes thermal expectation values as pure
state expectation values in an entangled state of two copies of the
original system.  The idea is to consider the state

$$| \Psi_{\beta} \rangle = \sum_n e^{- \frac{\beta E_n}{2}} |n
\rangle_1 \otimes |n \rangle_2 , $$ for a system whose Hamiltonian
is $$H = H_1 - H_2 .$$ This is a zero energy state, and we can
consider ordinary expectation values in this state of operators {\it
e.g.} that operate only in system 1.  It's easy to see that these
are just the thermal expectation values exhibited above.

Israel\cite{israel} pointed out that the Hawking-Hartle state of QFT
on the Kruskal extension of a black hole was precisely such a
Thermo-field state, and that this was an explanation for the bizarre
fact that time runs in opposite directions in the two branches of
the manifold. Maldacena\cite{malda} has recently revived interest in
this interpretation in connection with AdS black holes and
\cite{susskindetal} has extended Israel's result to the Bunch-Davies
vacuum of dS space.

All of this is straightforward and well known.  There is another
class of solutions with bifurcate horizons that has been
investigated in recent literature. These are the Guth-Farhi
solutions\cite{gf} and similar solutions in AdS
space\cite{shenkeretal}.  There are two characteristics of these
solutions that are different from either eternal black holes or dS
space.   They evolve from smooth initial data on a space-like slice.
The future asymptotic region looks like a black hole embedded in
either asymptotically flat or AdS space.   The singular region of
the black hole has a second horizon, the exterior region of which is
a dS space with a black hole embedded in it.

Since they evolve from smooth initial data, one might have
interpreted solutions like these as demonstrations that one can
``create a universe in the laboratory".  There is an implicit
violation of unitarity in this phrase.  It is clear that the the
observer in the asymptotically flat or AdS region can only see an
impure density matrix in the asymptotic future, because its state is
entangled with the state in the dS region.  Furthermore, the entropy
of the dS universe is {\it not} bounded by the entropy of the black
hole in the asymptotically flat or AdS region.

Fortunately, we do not have to assume that the smooth initial
conditions, which lead to this bifurcated future infinity, define a
quantum state in our theory.  Guth and Farhi proved that all such
solutions have a singularity at finite proper time in their past,
and their paper is entitled ``An obstacle to creating a universe in
the laboratory". In asymptotically flat space-time the observables
of a quantum theory of gravity are all encoded in a scattering
matrix. No scattering data on past infinity, can ever produce the
Guth-Farhi initial data on a space-like slice. Thus, despite its
smoothness and despite the fact that it falls off at infinity, this
solution has nothing to do with the scattering matrix of quantum
gravity in asymptotically flat space.

Our experience with eternal black holes and de Sitter space suggests
an alternate explanation of the meaning of this state in the quantum
theory.   Consider a density matrix $\rho$ for a quantum system
whose Hilbert space is ${\cal H}_1$.  Given any other quantum system
whose Hilbert space ${\cal H}_2$ has a dimension larger than or
equal to that of ${\cal H}_1$, we can construct $\rho$ by tracing
over ${\cal H}_2$ in some entangled state

$$|\Psi \rangle = \sum_{m,n} c_{mn} |\psi_m \rangle_1 |\psi_n
\rangle_2 . $$
$$\rho = \sum_{m,k,n} c^*_{mn} c_{kn} |\psi_k \rangle_1 \langle \psi_m
|_2 .$$ Neither ${\cal H}_2$ nor the choice of entangled state is
uniquely specified by $\rho$.  If $\rho$ has less than maximal
entropy in ${\cal H}_1$, then we can even make do with a lower
dimensional Hilbert space ${\cal H}_2$.   In quantum information
theory, constructions like this are used by adherents of the {\it
Church of the Larger Hilbert Space}(CLHS), and clever choices of the
entangled state can simplify arguments and calculations.  What is
clear is that no such construction implies in quantum mechanics that
the particular Hilbert space ${\cal H}_2$ or choice of the entangled
state, has anything to do with the quantum theory in ${\cal H}_1$.
Even those kneeling in the pews of the CLHS would laugh you out of
the room if you suggested as much.  An impure density matrix is a
statement of ignorance. That ignorance can be caused by the
correlation of the degrees of freedom we measure with any number of
other systems.  There is no way to tell which one.

So I would like to suggest that all solutions of Einstein's
equations with two causally disconnected future infinities, even if
they evolve from smooth initial data, be viewed as possibly
artificial ways of constructing a density matrix for one system by
entangling it with another.  Such solutions {\it do not} imply that
one of the quantum systems can be realized as a state in the other.
One simply cannot create a universe in the laboratory.

Indeed, as I have argued elsewhere\cite{isovac}, if I consider a
potential with an asymptotically flat solution and a dS solution,
and attempt to throw energy in from infinity in order to form a
bubble in which the field lies near the dS minimum over a sphere of
radius $R$, then before $R$ becomes as large as the dS radius, the
entire region is engulfed in a black hole.  This procedure cannot
realize a Guth-Farhi solution, because of the theorem of \cite{gf}.

Thus, if we restrict attention to solutions whose initial conditions
extrapolate smoothly to past infinity in Minkowski space, with small
localized perturbations corresponding to incoming wave packets, we
will never find a solution with a bifurcate horizon.  The Cosmic
Censorship conjecture is properly stated as the claim that the
future asymptotics of all such solutions corresponds to outgoing
wave packets, plus a finite number of finite area black holes. In
the quantum theory the black holes will decay and the S-matrix for
wave packets will be unitary.

A similar result may be conjectured for asymptotically AdS space.
Here we must be careful to impose the correct boundary conditions at
spatial infinity in order to be sure we are discussing solutions
that correspond to states in a given theory, rather than
perturbations of the original theory by a local operator on the
boundary\cite{horhert}.  The appropriate boundary conditions for
scattering in AdS space have been discussed by
\cite{polchsussgiddli}.



\end{document}